\documentclass[amsmath,prl, elsarticle, twocolumn] {revtex4}
\usepackage{bm}
\usepackage{enumerate}
\usepackage{graphicx}
\usepackage[dvips]{epsfig}
\usepackage{epsf}
\usepackage{xcolor}
\makeatletter

\newcommand{\Rmnum}[1]{\expandafter\@slowromancap\romannumeral #1@}

\makeatletter

\usepackage[T2A]{fontenc}
\usepackage[utf8]{inputenc}
\usepackage[russian]{babel}

\begin{document}
\title{De Haas-van Alphen effect and a giant temperature peak in heavy fermion SmB$_6$ compound}

\author{Vladimir A. Zyuzin}
\affiliation{L.D. Landau Institute for Theoretical Physics, 142432, Chernogolovka, Russia}
\begin{abstract}
In this paper we suggest a possible explanation of the giant temperature peak in the amplitude of the de Haas-van Alphen oscillations observed at very low temperatures in insulating SmB$_6$ system. Our theoretical model consists of two fermions with particle-like dispersion but with different masses, one much heavier than the other, which hybridize with each other to open up a gap at their degeneracy point. 
As a result of the hybridization a heavy-fermion hybrid appears at the Fermi level. Our results strongly suggest that it is exactly this heavy-fermion hybrid which results in the giant temperature peak. 
In addition we propose a scenario when this hybrid has edge states. 
\end{abstract}
\maketitle

Recent experiments where quantum oscillations in Kondo insulator SmB$_{6}$ were observed \cite{LiScience2014,SuchitraScience2015} still pose questions to the theory. 
The material is insulating due to a gapped Fermi surface of the conducting fermions. Nevertheless, is shows de Haas-van Alphen (dHvA) oscillations and the effect has deviations from the standard Lifshits-Kosevich \cite{LK,LP} dependence of the amplitude on temperature.

In our point of view, there is a number of main experimental facts drawn from \cite{LiScience2014,SuchitraScience2015} that a theory should address.
First of all, the observed frequency of dHvA oscillations of the insulator is that of the metallic phase of the system \cite{LiScience2014,SuchitraScience2015}, i.e. before it turned insulating via the gapping mechanism. 
In experiments it was proved by comparing frequencies of a material LaB$_{6}$ which has a similar band struture to SmB$_6$, but which never turns insulating.
Proposed theories \cite{KnolleCooperPRL15,TopodHvA2016,AlisultanovJETP2016,PalPRB2016,PalArxiv2022} do explain this fact in various models of insulators. 
Insulating state in these theories \cite{KnolleCooperPRL15,TopodHvA2016,AlisultanovJETP2016,PalPRB2016,PalArxiv2022} is formed due to the hybridization of electron-like and hole-like fermions at their degeneracy points, and as a consequence opening of an energy gap in the spectrum of fermions.

Secondly, another main outstanding question is the giant temperature peak observed at small temperatures \cite{SuchitraScience2015} in the temperature dependence of the amplitude of the dHvA oscillations of SmB$_6$. This is a drastic deviation from the standard Lifshits-Kosevich formula \cite{LK,LP} of the amplitude and there is no theory so far which explains the peak.
In this paper we give a possible explanation of this giant temperature peak.
In addition, certain corollaries on the nature of SmB$_6$ system can be drawn from our theory. 
Then, in order to verify our explanation of the peak, based on the corollaries, we provide new experimental signatures to look for in the samples of experiments \cite{SuchitraScience2015}.

The author has obtained some original theoretical details of the temperature dependence of the dHvA oscillations in correlated insulators in \cite{Zyuzin2023}. 
There the occurrence of the hybridization of the dispersive band corresponding to itinerant $d-$ electrons with the localized flat band of $f-$ electrons leading to the insulating gap in the system was studied. 
The case when the hybridization gap is treated self-consistently was considered, which matches the approach of \cite{AlloccaCooper}. 
This is the generalization of the Keldysh-Kopaev \cite{KeldyshKopaev} model to the presence of the magnetic field.
In \cite{Zyuzin2023} it was confirmed that the dHvA oscillations are possible, in accord with \cite{KnolleCooperPRL15,TopodHvA2016,AlisultanovJETP2016,PalPRB2016,PalArxiv2022}, and was found that instead of a standard Lifshits-Kosevich formula \cite{LK,LP} for the amplitude there are quantum oscillations of the amplitude with inverse temperature which show characteristic local maxima and minima. 
In \cite{Zyuzin2023} analytical expressions of the temperature at which the first peak occurs and its height were derived.
However, this result, as well as those of \cite{KnolleCooperPRL15,TopodHvA2016,AlisultanovJETP2016,PalPRB2016,PalArxiv2022}, don't explain any giant temperature peak similar to the one in \cite{SuchitraScience2015} in the expression for the amplitude of dHvA oscillations.

\begin{figure}[h] 
\centerline{
\begin{tabular}{ccc}
\includegraphics[width=0.3 \columnwidth]{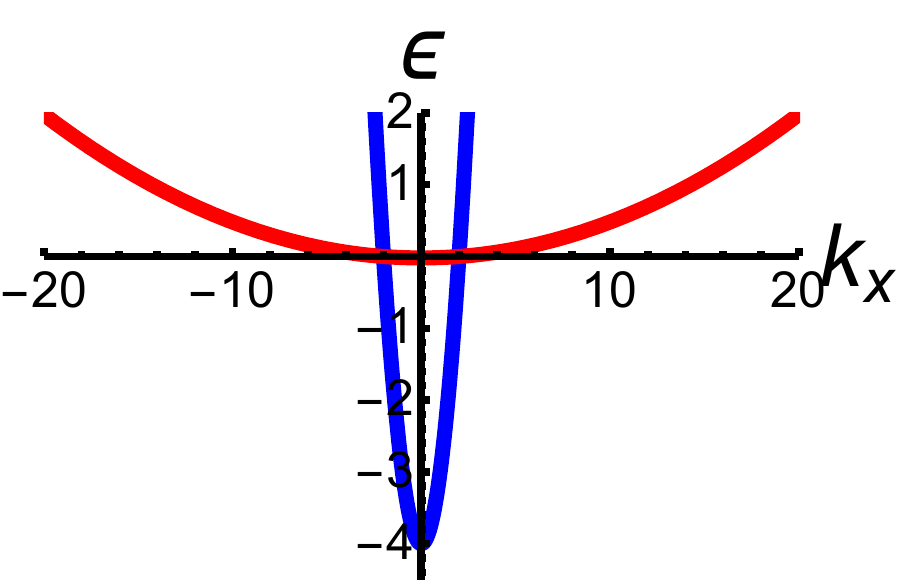}~~&
\includegraphics[width=0.3 \columnwidth]{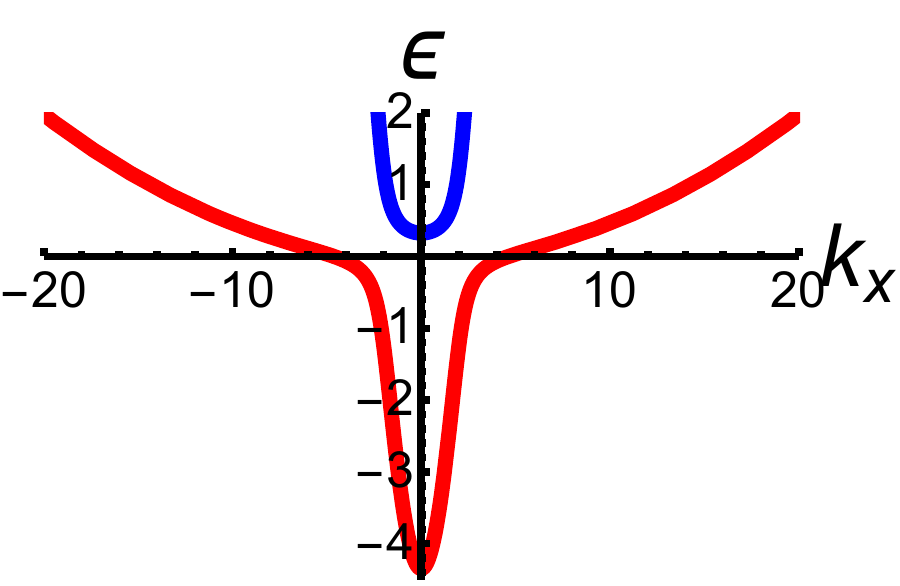}~~&
\includegraphics[width=0.3 \columnwidth]{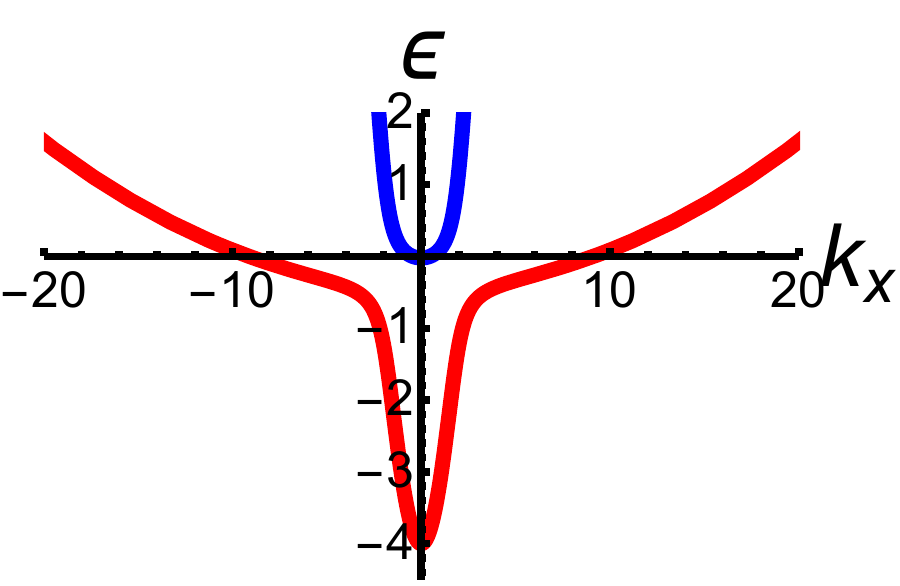}
\end{tabular}
}
\protect\caption{Schematics of a $k_{y}=0$ slice of the spectrum of the heavy fermion model given by the Hamiltonian Eq. (\ref{modelB}) and (\ref{modelC}) for $\alpha = 1/200$, $2m=1$, $\mu=4$ in corresponding units. We set the chemical potential to be equal $0$. 
Left: before the hybridization. Center: after the hybridization in the model Eq. (\ref{modelB}) for $\bar{\theta}\theta = 1.5$. Right: after the hybridization in the model Eq. (\ref{modelC}) for $\bar{\theta}\theta  = 1.5/\mu$. }

\label{fig:fig1}  

\end{figure}

\begin{figure}[h] 
\centerline{
\begin{tabular}{cc}
\includegraphics[width=0.5 \columnwidth]{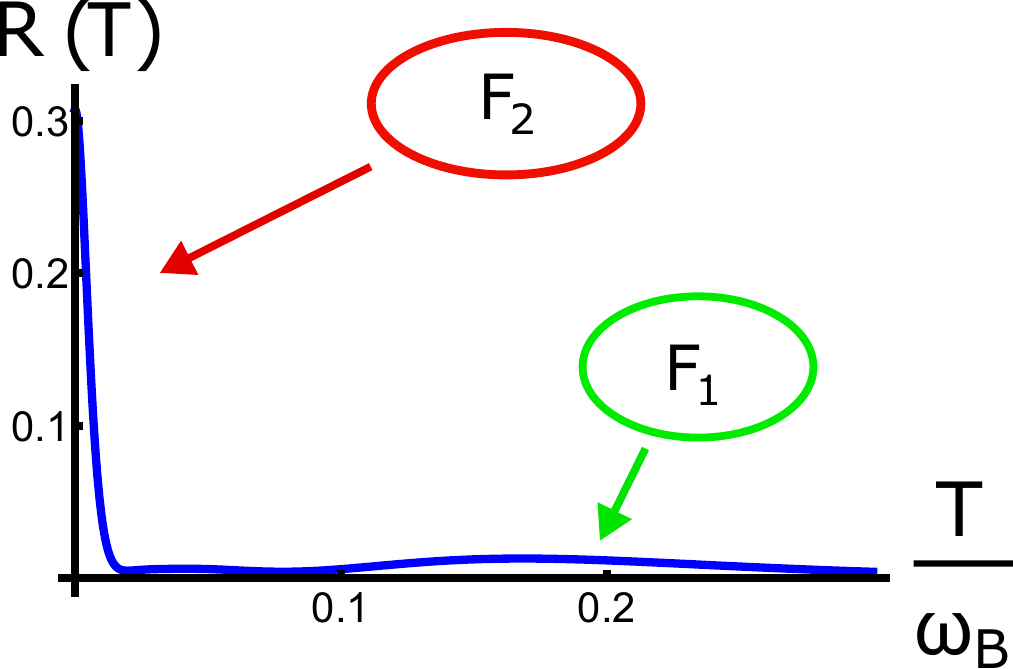}
\end{tabular}
}
\protect\caption{Temperature dependence of the amplitude of the dHvA oscillations of a system described by Eq. {\ref{modelB}}. For the illustration we chose $B=35\mathrm{T}$, $2\theta_{0}=35\mathrm{K}$, $\alpha=0.025$. }

\label{fig:fig2}  

\end{figure}

We will generalize the theoretical model of \cite{Zyuzin2023} to the case when $f-$ electrons are not localized, but rather have electron-like dispersion with a large mass. 
The Hamiltonian in the basis of
$
\bar{\psi} = \left[ \bar{\phi}_{\mathrm{d}} ,~ \bar{\phi}_{\mathrm{f}} \right]
$, and the same for 
$\psi$, is given by
\begin{align}\label{modelB}
H_{\mathrm{B}} = \int_{\bf k} \bar{\psi}\left[\begin{array}{cc}\xi_{\bf k} & \theta \\ \bar{\theta} & \alpha\xi_{\bf k} \end{array} \right]\psi ,
\end{align}
where $\alpha>0$ and $\alpha \ll 1$ is a measure of how heavy the $f-$ electrons are, $\xi_{\bf k} = \frac{{\bf k}^2}{2m} - \mu$ is chosen to conveniently describe the crossing of the two electrons spectra, $m$ is the effective mass, and $\theta$ corresponds to the hybridization between $d-$ and $f-$ electrons and in principle should be considered self-consistently just like it is done in \cite{Zyuzin2023}. 
Note that $\mu$ isn't the Fermi energy, but rather a parameter in the quadratic expansion of the spectrum. 
The Fermi energy is set to the crossing point of the two electron spectra, i.e. $\epsilon_{\mathrm{F}} = 0$. 
We could have chosen $\epsilon_{\mathrm{F}} \neq 0$. 
It is expected that in the slight vicinity of $\epsilon_{\mathrm{F}} = 0$ this deviation will not change our analysis presented below. 
However, a significant deviation will cause spatial dependence of the mean field solution for the hybridization. 
We wish to avoid this case in this paper.
Resulting dispersion assuming mean field ansatz for the hybridization, $\theta = \mathrm{const}(t;{\bf r})$ and $\theta^{*} = \bar{\theta}$, is 
\begin{align} \label{spectrumB}
\epsilon_{{\bf k};\pm} = \frac{1+\alpha}{2}\xi_{\bf k} \pm \sqrt{\left( \frac{1-\alpha}{2}\xi_{\bf k}  \right)^2 + \bar{\theta}\theta}.
\end{align}
A zero of the $\epsilon_{{\bf k};-}$ band is at $\xi_{\bf k} = \sqrt{\frac{\bar{\theta}\theta}{\alpha}}$, which is the Fermi surface of the heavy fermion $d-f$ hybrid. 
There is no zero of the $\epsilon_{{\bf k};+}$ fermion band. 
In Fig. (\ref{fig:fig1}) a $k_{y}=0$ slice of the spectrum Eq. (\ref{spectrumB}) before (left plot) and after (center plot) the hybridization is plotted. 
Let us now add a magnetic field and study the dHvA effect. Just like it is done in \cite{Zyuzin2023} we will only focus on the self-consistent equation on the hybridization, as it will already contain all necessary temperature dependence of the dHvA effect. The equation is obtained by first performing Hubbard-Stratonovich transformation which decouples interaction $H_{\mathrm{int}} = U \int_{x} \bar{\phi}_{\mathrm{d}}(x)\phi_{\mathrm{d}}(x) \bar{\phi}_{\mathrm{f}}(x)\phi_{\mathrm{f}}(x)$ between different fermions. As a result of the decoupling, new bosonic fields $\theta(x)$ and $\bar{\theta}(x)$ appear in the action (please see Eq. (1)). Finally, one has to vary the action with respect to this bosonic field in order to obtain the self-consistency equation for the mean field ansatz $\bar{\theta}(x) = \theta^{*}(x) = \mathrm{const}(x)$. We consider a model repulsive interaction $U >0$ which doesn't depend on coordinates, but in general it may be a function of coordinates.
In the magnetic field $B$ the self-consisten equation on the hybridization $\theta$ is
\begin{align}\label{equation}
\theta =  \theta \nu U \omega_{\mathrm{B}} \sum_{n} 
\frac{{\cal F}_{\epsilon_{n,+}} 
-
{\cal F}_{\epsilon_{n,-}} }{\sqrt{(1-\alpha)^2(\omega_{\mathrm{B}}n+\frac{\omega_{\mathrm{B}}}{2} - \mu)^2 +4\bar{\theta}\theta}},
\end{align}
where ${\cal F}_{x} = \tanh\left(\frac{x}{2T}\right) = 1-2n_{\mathrm{F}}(x)$, where $n_{\mathrm{F}}(x)$ is the Fermi-Dirac distribution function, and $\epsilon_{n;\pm}$ are obtained from Eq. (\ref{spectrumB}) by $\xi_{\bf k} \rightarrow \omega_{B} \left(n+\frac{1}{2}\right) - \mu$, where $n$ is the Landau level index and where $\omega_{B}$ is the cyclotron frequency $\omega_{B} = \frac{eB}{mc}$, and $\nu = \frac{m}{2\pi}$.
We again consider the two-dimensional case, while all of the results can be generalized to the three-dimensions straightforwardly. We utilize the Poisson summation formula to sum up over the Landau levels, 
\begin{align}\label{Poisson}
\sum_{n=0}g(n) = \int_{0}^{\infty}g(x) dx + \sum_{p\neq 0}\int_{0}^{\infty} e^{i2\pi p x} g(x) dx,
\end{align}
where $g(n)$ is some function. Equation Eq. (\ref{equation}) is then rewritten as $1=\nu U \omega_{\mathrm{B}}\sum_{p} {\cal R}_{p}$.
Details of the dHvA effect can be carried out from the $p=\pm 1$ terms of the sum. 
By following the lines and notations of \cite{Zyuzin2023}, the term with $p=1$ ($p=-1$ term is a complex conjugate of the $p=1$) reads as
\begin{align}\label{R+1}
{\cal R}_{+1} =
\frac{e^{i2\pi \frac{\mu}{\omega_{\mathrm{B}}}}}{(1-\alpha)\omega_{\mathrm{B}}}\int_{-\frac{\mu}{\omega_{\mathrm{B}}}+\frac{1}{2}}^{\frac{\Lambda}{\omega_{\mathrm{B}}}} e^{i2\pi x}
\frac{{\cal F}_{\epsilon_{x,+}}- {\cal F}_{\epsilon_{x,-}}}{ \sqrt{ x^2 +b^2 }} dx
\end{align}
where it is important to keep the lower limit as it is. Typically the limits at this stage are set to $\pm \infty$.  
The term with $p=-1$ is obtained by complex conjugation of Eq. (\ref{R+1}), i.e. ${\cal R}_{-1} = {\cal R}_{+1}^{*}$. 
We outline the distribution function
\begin{align}
{\cal F}_{\epsilon_{x,\pm}} = \tanh\left[\frac{\beta x \pm \sqrt{ x^2 +b^2 }}{4T/(1-\alpha)}\right],
\end{align}
where $\beta = \frac{1+\alpha}{1-\alpha}$ and $b = \frac{2\sqrt{\bar{\theta}\theta}}{(1-\alpha)\omega_{\mathrm{B}}}$.
We will now analyze the residues of the distribution function as a function of temperature.
The residues of the distribution function are 
\begin{align}\label{residues}
x_{m;\pm} 
= \frac{i}{\beta^2 - 1} \left[ \beta T_{m} \pm \sqrt{ T_{m}^2 - \left( \beta^2 - 1\right)b^2} \right],
\end{align}
where $T_{m} = \frac{4\pi T}{(1-\alpha) \omega_{\mathrm{B}}} (2m+1)$ is the dimensionless fermionic Matsubara frequency.
When $\beta = 1$, i.e. $\alpha = 0$, only $x_{m;-}$ works and we recover the case of \cite{Zyuzin2023}, indeed
\begin{align}
\lim_{\beta \rightarrow 1} x_{m;-}  = \frac{i}{2T_{m}}(T_{m}^2 + b^2).
\end{align}
This is also the case for $\alpha >0$ and $\alpha \ll 1$ at large tempeartures, when $T_{0}^2 \gg \left( \beta^2 - 1\right)b^2$, and we expect that the results of \cite{KnolleCooperPRL15,Zyuzin2023} hold at these temperatures. The frequency of the dHvA oscillations at these temperatures is $F_{1} = \mu$, consistent with previous works \cite{KnolleCooperPRL15,TopodHvA2016,AlisultanovJETP2016,PalPRB2016,PalArxiv2022}.
In the $T \rightarrow 0$ limit, we simply put $T=0$ under the square root in Eq. (\ref{residues}), 
\begin{align}
x_{m;\pm} 
= \mp \sqrt{ \frac{\bar{\theta}\theta}{\vert \alpha \vert} } \frac{1}{\omega_{\mathrm{B}}}  - i 0 \mathrm{sign}(2m+1),
\end{align}
and the oscillation of the heavy fermion $d-f$ hybrid becomes coherent in this limit.
A real part of the residue corresponds to the frequency of the oscillation with inverse magnetic field. 
The result in this limit is similar to the standard Lifshits-Kosevich formula \cite{LK,LP}. 
Following the lines of Ref. \cite{Zyuzin2023} we plot the amplitude of the oscillating part of the hybridization $\theta$ as a function of temperature in Fig. (\ref{fig:fig2}).
Oscillating part of the magnetization will have the same temperature dependence as shown in Fig. (\ref{fig:fig2}).

Now it became clear why the limits in the Eq. (\ref{R+1}) were kept fixed. 
When $\mu > \sqrt{ \frac{\bar{\theta}\theta}{\vert \alpha \vert} }$ both $x_{m;\pm}$ contribute to the integration, and it was safe to set 
limits to $\pm \infty$ in Eq. (\ref{R+1}) from the very beginning. This situation corresponds to the case when both $\epsilon_{{\bf k};\pm}$ are occupied.
However, we are interested in the opposite limit $\mu \leq \sqrt{ \frac{\bar{\theta}\theta}{\vert \alpha \vert} }$, i.e. in the limit of a heavy fermion with $\alpha \ll 1$ when only $\epsilon_{{\bf k};-}$ is occupied.
Then due to the fixed limits in Eq. (\ref{R+1}) only the $x_{m;-}$ residues contribute. The frequency $F_{2}$ of the dHvA oscillations at these low temperatures will correspond to the heavy-fermion $d-f$ hybrid's Fermi surface, namely $F_{2} = \mu +  \sqrt{ \frac{\bar{\theta}\theta}{\vert \alpha \vert} }$.

Let us connect our theoretical analysis with the experiment \cite{SuchitraScience2015}. 
In Fig. 4 of the \cite{SuchitraScience2015} a giant temperature peak in the temperature dependence of the amplitude of the dHvA in SmB$_{6}$ observed at very low temperatures is shown. A peak on its own is not that surprising as even regular Lifshits-Kosevich expression \cite{LK,LP} can give such a peak for fermions with a heavy mass. What is unusual in Fig. 4 of \cite{SuchitraScience2015} is a non-zero amplitude at higher temperatures. Conventional Lifshits-Kosevich expression for fermions with a heavy mass predicts an exponential suppression of the amplitude at high temperatures. The whole temperature range of Fig. 4 of the \cite{SuchitraScience2015} is plotted for a frequency of $330\mathrm{T}$ (in Teslas).

Our theory strongly suggests that the giant temperature peak in the amplitude of dHvA oscillations observed in \cite{SuchitraScience2015} is due to a heavy fermion which sets in at very small temperatures. The fact that the system is insulating suggests that all light mass fermions are gapped at the Fermi energy and 
they oscillate in the dHvA effect with a suppressed amplitude which stretches to high temperatures (just like it is shown in Fig. (\ref{fig:fig2})). The choice of the Hamiltonian Eq. (\ref{modelB}) which has both features is thus justified. 
However, in our theory, the high and low temperature regimes shown in Fig. (\ref{fig:fig2}) oscillate with different frequencies $F_{1} = \mu$ and $F_{2}=\mu+ \frac{\theta_{0}}{\sqrt{\alpha } }$ correspondingly while in the experiment \cite{SuchitraScience2015} the two regimes oscillate with the same frequency. 
There is no way they can be strictly the same in our model and we rely on coincidences in order to explain \cite{SuchitraScience2015}. We suggest two plausible scenarios. 
In the first one we claim that there is a coincidence and a second harmonic of oscillations with $F_{1}$ equals to $F_{2}$, in other words $2F_{1} \approx F_{2}$. This is still a heavy fermion limit defined by $\mu \leq \frac{\theta_{0}}{\sqrt{\alpha } }$ when only $\epsilon_{{\bf k};-}$ is occupied and our theoretical predictions hold. We remind that for $\mu > \frac{\theta_{0}}{\sqrt{\alpha } }$ both branches of spectrum will be occupied and the system will be metallic. Furthermore, we examine Fig 2a in \cite{SuchitraScience2015} and connect the largest peak corresponding to frequency of $F_{3}^{\mathrm{exp}}=330\mathrm{T}$ with $F_{2}$, while a peak at $F_{2}^{\mathrm{exp}}=170\mathrm{T}$ with $F_{1}$. Therefore, indeed $2F_{2}^{\mathrm{exp}} \approx F_{3}^{\mathrm{exp}}$. Our proposal can be realistic since that Ref. \cite{LiScience2014} experimentally observed substantial amplitude of the second and third harmonic of the dHvA oscillations (Fig 2b in Ref. \cite{LiScience2014}) of a given frequency in SmB$_{6}$.

In our second scenario we propose that there are two fermion pockets, labelled as $1$ and $2$, each described by Hamiltonian Eq. (\ref{modelB}), and both become hybridized at small temperatures.
The first pocket has $\alpha_{1}$, $\mu_{1}$, and $\theta_{1}$ parameters in the heavy fermion regime, while the second one $\alpha _{2}= 0$ (or $\alpha < 0$), $\mu_{2}$ and $\theta_{2}$, i.e. described by a model which has no heavy-fermion hybrid and thus has no giant peak. 
Then one may assume that somehow a $\mu _{1}+ \frac{\theta_{1}}{\sqrt{\alpha_{1}}}\approx \mu_{2}$ coincidence occurred such that oscillations at high temperatures in Fig. (\ref{fig:fig2}) are with frequency equal to $\mu_{2}$, while the giant peak comes from oscillations with a frequency equal to $\mu_{1}+\frac{\theta_{1}}{\sqrt{\alpha_{1}}}$. 
First three lowest frequencies shown in Fig. (2A) in \cite{SuchitraScience2015} suggest that this scenario might as well be correct. 
In particular, either $F_{1}^{\mathrm{exp}}= 50 \mathrm{T}$ or $F_{2}^{\mathrm{exp}}=170\mathrm{T}$ frequency stands for $\mu_{1}$, while $F_{3}^{\mathrm{exp}}=330\mathrm{T}$ stands for both $\mu_{1} + \frac{\theta_{1}}{\sqrt{\alpha_{1}}}$ and $\mu_{2}$.

Let us now check whether the quantum oscillations with inverse tempearture proposed in \cite{Zyuzin2023} do exist. We look at the Fig. (4) in \cite{SuchitraScience2015}, and associate a first bump occuring at $T^{*} \approx 8\mathrm{K}$ with the predicted in \cite{Zyuzin2023} position of the largest peak, $T_{\mathrm{peak}} = \frac{\theta_{0}}{\pi}$, then we extract the value of the hybridization gap to be $2\theta_{0} = 2\pi T^{*} \approx 48 \mathrm{K}$ which is consistent with the energy gap of $40 \mathrm{K}$ observed from activated electrical conductivity behaviour \cite{SuchitraScience2015}. 
Furthermore, with that value and using the $2F_{1}=F_{2}$ condition which reads as $\mu = \frac{\theta_{0}}{\sqrt{\alpha}}$ we can obtain the flatness of the $f-$ fermions in our model to be $\alpha = \frac{\theta_{0}^2}{\mu^2} = \frac{24^2}{330^2}\approx 0.005 = \frac{1}{200}$.
Finally, we can estimate the temperature $T_{\mathrm{g}}$ at which the giant temperature peak starts to grow by setting $ T_{m=0}^2 - \left( \beta^2 - 1\right)b^2 =0 $ which reads as $T=\frac{\sqrt{\alpha}}{\pi}\theta_{0}$, and get $T_{\mathrm{g}} \approx 0.5 \mathrm{K}$. This number is in good agreement with the Fig. 4 in \cite{SuchitraScience2015}.

The region in which quantum oscillations as a function of temperature were predicted by the author in \cite{Zyuzin2023} needs to be scanned in details to confirm that the small heights observed in Fig. 4 of \cite{SuchitraScience2015} at $T\approx 8\mathrm{K}$ and $T\approx 5\mathrm{K}$ are not just the deviations within the uncertainties but rather unique physical features. We stress that these oscillations are guaranteed if the giant low temperature peak is observed.

We claim that SmB$_6$ samples in experiments \cite{SuchitraScience2015} are not insulating, but rather systems with a heavy fermion at the Fermi energy. Although the hybridization between itinerant $d-$ electrons and localized $f-$ electrons happens at the transition temperature and the system seemingly establishes a collective gap for the $d-$ electrons at the Fermi energy, a possible slight electron-like dispersion of the $f-$ electrons results in a large Fermi surface of the heavy fermion $d-f$ hybrid which is never gapped and is thus metallic. It is quite likely that the observed saturation of the resistivity in SmB$_6$ at very low temperature is due to the residual resistivity of this heavy fermion $d-f$ hybrid. If so, it rules out the proposal \cite{DzeroSunGalitskiColeman} that SmB$_6$ is a topological insulator.
Therefore, since our theory proposes an existence of the $d-f$ heavy fermion hybrid at the Fermi level, which becomes visible in dHvA oscillations at very low temperatures,
we speculate that Shubnikov-de Haas effect with $F_{2} = \mu +  \sqrt{ \frac{\bar{\theta}\theta}{\vert \alpha \vert} }$ frequency can be observed at these very low temperatures despite large resistivity.

We note that there exists a theoretical proposal \cite{PalPRB2019} which claims that SmB$_6$ might be an insulator-like metal. Ref. \cite{PalPRB2019} studied Hamiltonian which is Eq. (\ref{modelB}) but with $\alpha=0$ and with the Fermi energy set to the bottom of the conduction band (slight doping). A model proposed in our paper is always metallic with a heavy fermion $d-f$ hybrid at the Fermi energy.

Let us now demonstrate how the system may be metallic, not a topological insulator, but nevertheless have edge states.
Our choice of a theoretical model is motivated by original works on the edge states in metallic systems \cite{DyakonovKhaetskii1981,StanescuGalitski2006,ZyuzinSilvestrovMishchenko2007} and topological insulators \cite{SovietTI1,SovietTI2}.
For example, from \cite{ZyuzinSilvestrovMishchenko2007} we know that there will be edge states in two-dimensional electron gas with Rashba spin-orbit coupling.
To map the system in \cite{ZyuzinSilvestrovMishchenko2007} on to our system we must require the hybridization gap to be odd in momentum.
Suppose we have found such an interaction between $d-$ and $f-$ fermions which stabilizes a hybridization gap with $\theta_{\bf k}  = - \theta_{-{\bf k}}$ symmetry. 
Then our minimal model Hamiltonian is 
\begin{align}\label{modelC}
H_{\mathrm{C}}  = \int_{\bf r} \bar{\psi}({\bf r})\left[\begin{array}{cc}\hat{\xi}_{\bf k} & \theta (\hat{k}_{x} + i\hat{k}_{y}) \\ \bar{\theta} (\hat{k}_{x} - i\hat{k}_{y}) & \alpha \hat{\xi}_{\bf k} \end{array} \right]\psi({\bf r}) ,
\end{align}
where $\theta_{\bf k} = \theta (\hat{k}_{x} + i\hat{k}_{y})$, hats denote operators, and the basis now contains spin $\bar{\psi}_{{\bf k}} =  \left( \bar{\psi}_{\mathrm{d};\uparrow},~ \bar{\psi}_{\mathrm{f};\downarrow} \right)$ (there is an independent and similar Hamiltonian for opposite spins basis). We note that for $\alpha = -1$, the model corresponds to the Volkov-Pankratov model of topological insulators \cite{SovietTI1,SovietTI2}, whereas $\alpha =1$ corresponds to the two dimensional electron gas with Rashba spin-orbit coupling, whose edge states were studied in \cite{ZyuzinSilvestrovMishchenko2007}. The spectrum of the bulk fermions is
\begin{align} \label{spectrumC}
\epsilon_{{\bf k};\pm} = \frac{1+\alpha}{2}\xi_{\bf k} \pm \sqrt{\left( \frac{1-\alpha}{2}\xi_{\bf k}  \right)^2 + \bar{\theta}\theta k^2},
\end{align}
which we show in the right plot of Fig. (\ref{fig:fig1}).
If the chemical potential is set to zero, then both $\epsilon_{{\bf k};\pm}$ bands are occupied. 
In the right plot of the Fig. (\ref{fig:fig1}) a small pocket (in blue) corresponds to $\epsilon_{{\bf k};+}$ band. The spinors are
\begin{align}
\psi_{\pm;{\bf k}} = \frac{1}{\sqrt{2}} \left[ \begin{array}{c} \pm \frac{\sin\zeta_{\bf k}}{\sqrt{1\pm\cos\zeta_{\bf k}}}e^{i\chi_{\bf k}} \\ \sqrt{1\pm\cos\zeta_{\bf k}} \end{array}\right],
\end{align}
where $\chi_{\bf k} = \mathrm{arctan}\left( \frac{k_{y}}{k_{x}} \right)$, and which is odd in either $k_{x}$ or $k_{y}$, and $\cos(\zeta_{\bf k}) = \frac{1-\alpha}{2}\xi_{\bf k}\left[\sqrt{\left( \frac{1-\alpha}{2}\xi_{\bf k}  \right)^2 + \bar{\theta}\theta k^2}\right]^{-1}$. Let us set a hard wall boundary at $x=0$ and assume a free space at $x>0$, then $\psi(x=0,y) = 0$. Then the $+$ and $-$ spinors mix in order to satisfy the boundary condition, because $\chi_{k_{x};k_{y}} = - \chi_{-k_{x};k_{y}}$. 
We define
\begin{align}
\zeta =  1- \frac{2}{1+ \sqrt{1+\frac{2\alpha \mu}{m \bar{\theta}\theta}}}   > 0,
\end{align}
which appears as a parameter in our analysis.
It can be shown that for $k_{y}^2 < \zeta 2m\mu$ there are no edge states. Plane waves from both $\epsilon_{{\bf k};\pm}$ bands scatter from the boundary.
When $k_{y}^2 > \zeta 2m\mu$ the edge states originating from $\epsilon_{{\bf k};+}$ band appear which decay from the boundary as $\psi_{+;k_{y}}({\bf r}) \propto e^{-\sqrt{k_{y}^2 - \zeta 2m \mu}x} e^{ik_{y}y}$. We note that the edge states will also be present if our minimal model Eq. (\ref{modelC}) will be generalized to more than two bands provided that oddness of the hybridization gap in momentum maintains.

To conclude, we have proposed a theoretical model described in Eq. (\ref{modelB}) to shed light on the giant temperature peak of the dHvA oscillations amplitude experimentally observed in SmB$_6$ at very low temperatures \cite{SuchitraScience2015}. Main conclusion derived from our model is that the system is never truly insulating but rather metallic with a heavy fermion $d-f$ hybrid at the Fermi energy, which might appear insulating due to its heavy mass. In the Fig. (\ref{fig:fig2}) we do obtain the giant temperature peak associated with the oscillations of this heavy fermion $d-f$ hybrid. The pitfall of the model is that, as shown in Fig. (\ref{fig:fig2}), the high temperature tail and the giant temperature peak correspond to oscillations with two different frequencies, $F_{1}$ and $F_{2}$ correspondingly. There is no way to make them the same. We suggested that either $F_{2}\approx 2F_{1}$ or there is a coincidence of frequencies from different Fermi pockets. There are experimental signatures in \cite{LiScience2014,SuchitraScience2015} in support of these two hypothesis. 
It has been debated whether SmB$_{6}$ is a topological insulator or not. We proposed a scenario described by the Hamiltonian Eq. (\ref{modelC}) when a heavy fermion $d-f$ hybrid being metallic has edge states.

\textit{Funding.}
The author is supported by the Foundation for the Advancement of Theoretical Physics and Mathematics BASIS.

\textit{Conflict of interestes.}
None.

\textit{Acknowledgements.}
The author thanks I.S. Burmistrov, A.M. Finkel'stein, M.M. Glazov, P.D. Grigoriev, A. Kamenev, D.G. Yakovlev, and A.Yu. Zyuzin for helpful discussions.
VAZ is grateful to Pirinem School of Theoretical Physics.

\begin{widetext}

\newpage
\clearpage
\onecolumngrid
\begin{center}
\rule{0.38\linewidth}{1pt}\\
\vspace{-0.37cm}\rule{0.49\linewidth}{1pt}
\end{center}
\setcounter{section}{0}
\setcounter{equation}{0}
\setcounter{figure}{0}

\section{Supplemental Material to  "De Haas-van Alphen effect and a giant temperature peak in heavy fermion SmB$_6$ compound"}

\subsection{Fermi surface of the heavy fermion $d-f$ hybrid}
Here we perform a simple excercise of finding Fermi surface of the heavy fermion $d-f$ hybrid.
The spectrum given in the Main Text is
\begin{align} \label{spectrumSM}
\epsilon_{{\bf k};\pm} = \frac{1+\alpha}{2}\xi_{\bf k} \pm \sqrt{\left( \frac{1-\alpha}{2}\xi_{\bf k}  \right)^2 + \bar{\theta}\theta},
\end{align}
where, we remind, $\xi_{\bf k} = \frac{{\bf k}^2}{2m} - \mu$.
Zero of the $\epsilon_{{\bf k};-} $ occurs only for $\xi_{\bf k} > 0$ and is found to be
\begin{align}
\xi_{\bf k} = \sqrt{\frac{\bar{\theta}\theta}{\alpha}},
\end{align}
which defines the Fermi energy of the heavy fermion $d-f$ hybrid,
\begin{align}
\epsilon_{\mathrm{F}}^{(-)} = \mu +  \sqrt{\frac{\bar{\theta}\theta}{\alpha}}.
\end{align}
This Fermi energy corresponds to the red curve in the right plot of Fig. (\ref{fig:fig1sm}) crossing zero (chemical potential is seto to zero). 
Let us check that there is no Fermi surface associated with the zeros of $\epsilon_{{\bf k};+}$ branch for the heavy fermion limit $\mu <\sqrt{\frac{\bar{\theta}\theta}{\alpha}}$. 
If any, it would occur only for $\xi_{\bf k} < 0$, then we solve $\epsilon_{{\bf k};+} =0$,
\begin{align}
\xi_{\bf k} = -\sqrt{\frac{\bar{\theta}\theta}{\alpha}},
\end{align}
and get
\begin{align}
\epsilon_{\mathrm{F}}^{(+)} \equiv \frac{(k^{(+)}_{\mathrm{F}})^2}{2m} = \mu -  \sqrt{\frac{\bar{\theta}\theta}{\alpha}},
\end{align}
and since we are working in the heavy fermion limit $\mu <\sqrt{\frac{\bar{\theta}\theta}{\alpha}}$,  there is no solution for the Fermi momentum.
It means the $\epsilon_{{\bf k};+}$ branch is unoccupied. This is indeed the case since only the red curve in the right plot of Fig. (\ref{fig:fig1sm}) corresponding to $\epsilon_{{\bf k};-} $ branch crosses zero.

\begin{figure}[h] 
\centerline{
\begin{tabular}{cc}
\includegraphics[width=0.4 \columnwidth]{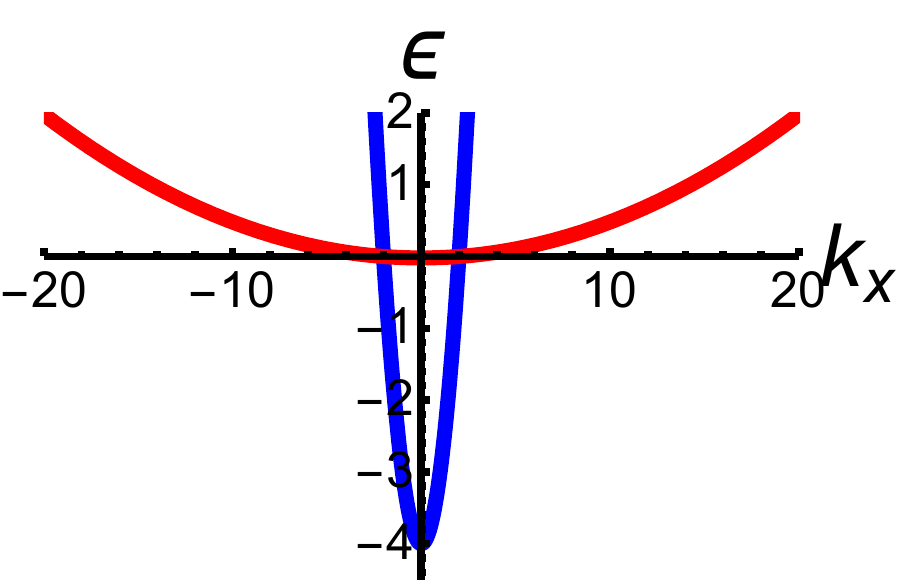}~~~~&
\includegraphics[width=0.4 \columnwidth]{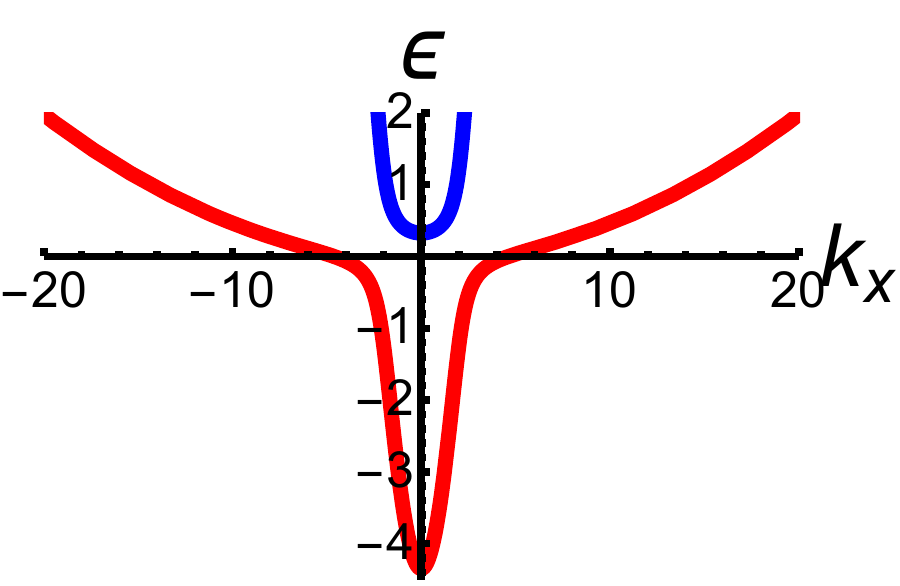}
\end{tabular}
}
\protect\caption{Schematics of a $k_{y}=0$ slice of the spectrum of the heavy fermion model given by Eq. (\ref{spectrumSM}) for $\alpha = 1/200$, $2m=1$, $\mu=4$, and $\bar{\theta}\theta = 1.5$ in corresponding units. We set the chemical potential to be equal $0$. The numerical values  (not directly related to the experimental values) of the parameters are chosen to only picture the Fermi surface after the hybridization equal to approximately $3$ times larger than that before. 
Left: before the hybridization. Right: after the hybridization.}

\label{fig:fig1sm}  

\end{figure}

In passing, we note that there are no edge states in this Hamiltonian for the hard-wall boundary potential.
Although, there is a spinor structure and the $\epsilon_{{\bf k};+} = 0$ allows for the edge states in the 
heavy fermion limit, there is no way for it to meet the boundary condition.
This is because the spinor structure of the $\epsilon_{{\bf k};-} = 0$ states is insensitive to $k_{x/y} \rightarrow -k_{x/y}$, such that the in-boundary-going and the out-boundary-going plane waves have the same spinor structure and satisfy the boundary condition together.
Finally, there are no edge states of the $\epsilon_{{\bf k};-} = 0$ spectrum to partner it up with the edge state of the $\epsilon_{{\bf k};+} = 0$ in order to meet the boundary condition.

\subsection{Summation over the Landau levels}
The non-linear equation reads as
\begin{align}
1 =  \nu U B \sum_{n} 
\frac{{\cal F}_{\epsilon_{n,+}} 
-
{\cal F}_{\epsilon_{n,-}} }{\sqrt{(1-\alpha)^2(\omega_{\mathrm{B}}n+\frac{\omega_{\mathrm{B}}}{2} - \mu)^2 +4\bar{\theta}\theta}},
\end{align}

Note that at $T=0$ expression
\begin{align}
\frac{1}{(1-\alpha)}
\int_{-\frac{\mu}{\omega_{\mathrm{B}}}+\frac{1}{2}}^{\sqrt{\frac{\theta^2}{\alpha}}} e^{i2\pi x}
\frac{2}{ \sqrt{ x^2 +\frac{4\bar{\theta}\theta}{(1-\alpha)^2\omega_{\mathrm{B}}^2} }} dx
\end{align}
has the upper limit $\propto \theta$, then there is a possibility of the strong coupling situation where the mean field solution does not exist.

\begin{align}
&
\frac{e^{i2\pi \frac{\mu}{\omega_{\mathrm{B}}}}}{(1-\alpha)}\int_{-\frac{\mu}{\omega_{\mathrm{B}}}+\frac{1}{2}}^{\frac{\Lambda}{\omega_{\mathrm{B}}}} e^{i2\pi x}
\frac{{\cal F}_{\epsilon_{x,+}}- {\cal F}_{\epsilon_{x,-}}}{ \sqrt{ x^2 +\frac{4\bar{\theta}\theta}{(1-\alpha)^2\omega_{\mathrm{B}}^2} }} dx
\\
=
&
-\frac{e^{i2\pi \frac{\mu}{\omega_{\mathrm{B}}}}}{(1-\alpha)}\int_{-\frac{\Lambda}{\omega_{\mathrm{B}}} }^{\frac{\mu}{\omega_{\mathrm{B}}}-\frac{1}{2}} e^{-i2\pi x}
\frac{{\cal F}_{\epsilon_{x,-}}}{ \sqrt{ x^2 +\frac{4\bar{\theta}\theta}{(1-\alpha)^2\omega_{\mathrm{B}}^2} }} dx
-
\frac{e^{i2\pi \frac{\mu}{\omega_{\mathrm{B}}}}}{(1-\alpha)}\int_{-\frac{\mu}{\omega_{\mathrm{B}}}+\frac{1}{2}}^{\frac{\Lambda}{\omega_{\mathrm{B}}}} e^{i2\pi x}
\frac{{\cal F}_{\epsilon_{x,-}}}{ \sqrt{ x^2 +\frac{4\bar{\theta}\theta}{(1-\alpha)^2\omega_{\mathrm{B}}^2} }} dx
\end{align}

\subsection{A model of the heavy fermion $d-f$ hybrid with edge states}
We generalize our model of heavy fermion $d-f$ hybrid to the case when the hybridization between the $d-$ and $f-$ fermions is of the $p-$wave type.
In another words, the model Hamiltonian is
\begin{align}\label{modelC_SM}
\hat{H}_{\mathrm{C}}  = \int_{\bf r} \bar{\psi}({\bf r})\left[\begin{array}{cc}\hat{\xi}_{\bf k} & \theta (\hat{k}_{x} + i\hat{k}_{y}) \\ \bar{\theta} (\hat{k}_{x} - i\hat{k}_{y}) & \alpha \hat{\xi}_{\bf k} \end{array} \right]\psi({\bf r}) ,
\end{align}
where again $\alpha >0$ and $\alpha \ll 1$, and $\theta$ and $\bar{\theta}$ are constants defining the hyridization.
In order to meet the $p-$wave structure, one needs to involve spins of the fermions.
The spinors are labelled as
\begin{align}
\psi_{1;{\bf k}} =  \left[ \begin{array}{c} \psi_{\mathrm{d};\uparrow} \\ \psi_{\mathrm{f};\downarrow} \end{array}\right],
\end{align}
where $\uparrow$ and $\downarrow$ are spin-up and spin-down states correspondingly.
There is an independent Hamiltonian for the 
\begin{align}
\psi_{2;{\bf k}} =  \left[ \begin{array}{c} \psi_{\mathrm{d};\downarrow} \\ \psi_{\mathrm{f};\uparrow} \end{array}\right],
\end{align}
spinor, which is obtained from Eq. (\ref{modelC_SM}) by $k_{y}\rightarrow -k_{y}$ replacement, and thus is studied in the same way as the Eq. (\ref{modelC_SM}). Indeed, the two blocks don't mix with each other, and can be studied separately. Finally, we point out that if $\alpha <0$ then the Hamiltonian descibes a Volkov-Pankratov model of topological insulators. We remind that we study the $\alpha > 0$ and $\alpha \ll 1$ case.
In what follows we omit the spinor index $1$ or $2$, and focus only on the $1$ block.
The dispersion of Eq. (\ref{modelC_SM}) is found to be
\begin{align} \label{spectrum_SM_C}
\epsilon_{{\bf k};\pm} = \frac{1+\alpha}{2}\xi_{\bf k} \pm \sqrt{\left( \frac{1-\alpha}{2}\xi_{\bf k}  \right)^2 + \bar{\theta}\theta k^2}.
\end{align}
\begin{figure}[h] 
\centerline{
\centerline{
\begin{tabular}{cc}
\includegraphics[width=0.4 \columnwidth]{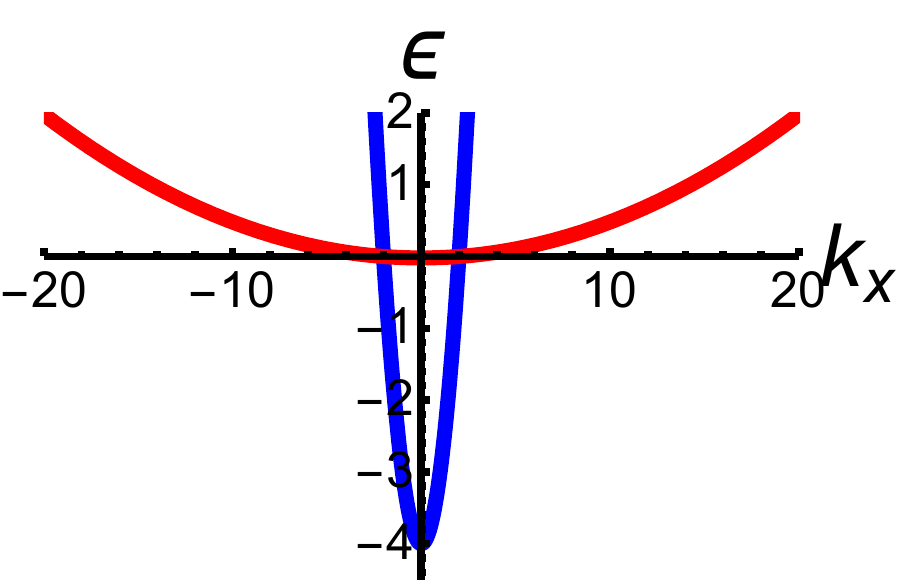}~~~~&
\includegraphics[width=0.4 \columnwidth]{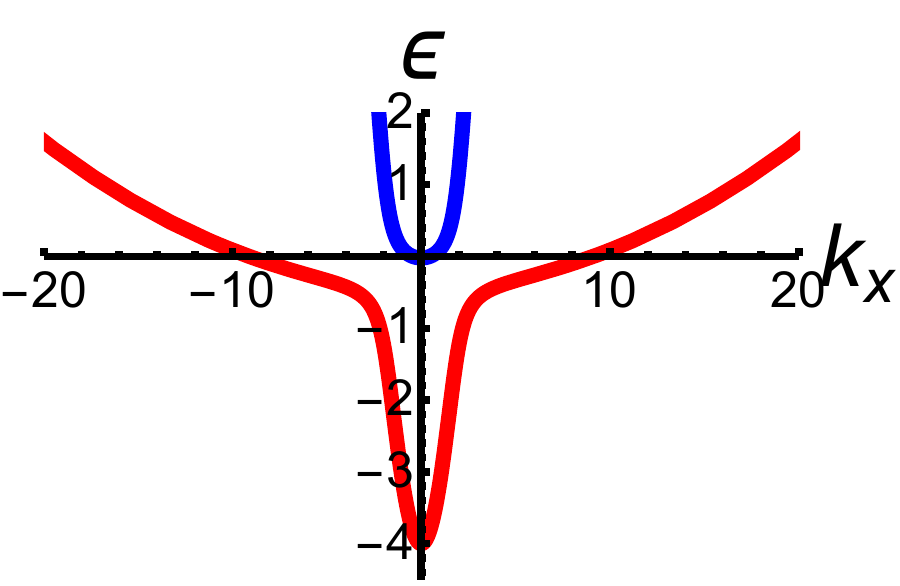}
\end{tabular}
}
}
\protect\caption{Schematics of a $k_{y}=0$ slice of the spectrum of the heavy fermion model given by the Hamiltonian Eq. (\ref{modelC}) for $\alpha = 1/200$, $2m=1$, $\mu=4$, and $\bar{\theta}\theta = 1.5/\mu$ in corresponding units. Left: before, right: after the hybridization.}

\label{fig:fig2sm}  

\end{figure}

\begin{align}
\psi_{\pm;{\bf k}} = \frac{1}{\sqrt{2}} \left[ \begin{array}{c} \pm \frac{\sin\zeta_{\bf k}}{\sqrt{1\pm\cos\zeta_{\bf k}}}e^{i\chi_{\bf k}} \\ \sqrt{1\pm\cos\zeta_{\bf k}} \end{array}\right]
\end{align}

It is clear we can't satisfy the boundary condition only with fermions from $\epsilon_{{\bf k};-}$, indeed
\begin{align}
C \psi_{-;k_{x},k_{y}}  + \psi_{-;-k_{x},k_{y}} \neq 0,
\end{align}
where as usual spinor with $-k_{x}$ is the incoming to the boundary, while the one with $k_{x}$ is the outgoing, and where $C$ is some coefficient to be found.
This is because $e^{i\chi_{-k_{x},k_{y}}} = e^{-i\chi_{k_{x},k_{y}}}$ and we can't satisfy two different equations with one parameter $C$, namely
\begin{align}
&
Ce^{-i\chi_{k_{x},k_{y}}} + e^{i\chi_{k_{x},k_{y}}} = 0, \\
&
C +1 = 0.
\end{align}
In order to satisfy boundary condition, we need to involve wave function corresponding to the $\epsilon_{+}$ band.
To see how it appears we utilize the method of Laplace transformation. 
We define
\begin{align}
\psi(q) = \int_{0}^{\infty} \psi(x)e^{-qx} dx.
\end{align}
Then
\begin{align}
&
\int_{0}^{\infty} \left[ \partial_{x}^2 \psi(x)\right] e^{-qx} dx = -\left. \psi^{\prime}(x)\right|_{x=0} + q^2 \psi(q), 
\\
&
\int_{0}^{\infty} \left[ \partial_{x} \psi(x)\right] e^{-qx} dx = q \psi(q).
\end{align}
We will use $\psi^{\prime}(0) \equiv \left. \psi^{\prime}(x)\right|_{x=0}$ definition for brevity.
Then the $\psi(q)$ is obtained from
\begin{align}
2m\int_{0}^{\infty}e^{-qx} \left( \hat{H} -  E \right)\psi(x) dx &= 
\left[\begin{array}{cc} -q^2 -   2m \mu_{k_{y}}- 2mE & 2m\theta \left( -iq + ik_{y} \right) \\ 2m\bar{\theta} \left( -iq - ik_{y} \right) & \alpha(-q^2 - 2m \mu_{k_{y}}) - 2mE   \end{array}\right]\psi(q)+ \left[\begin{array}{cc} 1 & 0 \\ 0 & \alpha \end{array} \right]\psi^{\prime}(0) 
\\
&= 0,
\end{align}
where $2m\mu_{k_{y}} = 2m\mu - k_{y}^2 $. Therefore, 
\begin{align}
\psi(q) = - \left[\begin{array}{cc} -q^2 -   2m \mu_{k_{y}}- 2mE & 2m\theta \left( -iq + ik_{y} \right) \\ 2m\bar{\theta} \left( -iq - ik_{y} \right) & \alpha(-q^2 - 2m \mu_{k_{y}}) - 2mE   \end{array}\right]^{-1}   \left[\begin{array}{cc} 1 & 0 \\ 0 & \alpha \end{array} \right]\psi^{\prime}(0). 
\end{align}
The inverse is 
\begin{align}
\left[\begin{array}{cc} -q^2 -   2m \mu_{k_{y}}- 2mE & 2m\theta \left( -iq + ik_{y} \right) \\ 2m\bar{\theta} \left( -iq - ik_{y} \right) & \alpha(-q^2 - 2m \mu_{k_{y}}) - 2mE   \end{array}\right]^{-1} = \frac{1}{det}
\left[\begin{array}{cc}   \alpha(-q^2 - 2m \mu_{k_{y}}) - 2mE   & -2m\theta \left( -iq + ik_{y} \right) \\ -2m\bar{\theta} \left( -iq - ik_{y} \right) &  -q^2 -   2m \mu_{k_{y}}- 2mE  \end{array}\right],
\end{align}
where $det$ is the determinant
\begin{align}
det = \left[   -q^2 -   2m \mu_{k_{y}}- 2mE \right] \left[ \alpha(-q^2 - 2m \mu_{k_{y}}) - 2mE  \right] - (2m)^2\bar{\theta}\theta (k_{y}^2 - q^2).
\end{align}
The residues are defined by the $det = 0$ equation.
We are interested in $E=0$ state. We define $Q^2 \equiv q^2 + 2m\mu_{k_{y}}$, then $det = 0$ reads
\begin{align}
\alpha Q^4 + (2m)^2 \bar{\theta}\theta Q^2 - (2m)^2\bar{\theta}\theta 2m\mu = 0.
\end{align}
Solutions are
\begin{align}
Q^2_{\pm} = \frac{(2m)^2 \bar{\theta}\theta}{2\alpha}\left[ -1 \pm \sqrt{1+\frac{8\alpha m\mu}{(2m)^2 \bar{\theta}\theta}}\right]. 
\end{align}
The $-$ solution is always negative for $2m\mu > k_{y}^2$, and it describs two plane waves (in-going to the boundary and out-going from the boundary) corresponding to the only occupied with fermions $\epsilon_{-;{\bf k}}$ band. 
The $+$ solution can rewritten to a more convenient form,
\begin{align}
Q^{2} = \frac{2}{1+ \sqrt{1+\frac{8\alpha m\mu}{(2m)^2 \bar{\theta}\theta}}} 2m\mu,
\end{align}
therefore,
\begin{align}
q^2 - k_{y}^2 = \left( \frac{2}{1+ \sqrt{1+\frac{8\alpha m\mu}{(2m)^2 \bar{\theta}\theta}}}  - 1 \right) 2m\mu \equiv -\zeta 2m\mu,
\end{align}
where 
\begin{align}
\zeta =  1- \frac{2}{1+ \sqrt{1+\frac{8\alpha m\mu}{(2m)^2 \bar{\theta}\theta}}}   > 0.
\end{align}
Therefore, for $\zeta 2m \mu >k_{y}^2 > 0$ there are two plane waves (in-going to the boundary and out-going from the boundary) corresponding to a small pocket of the $\epsilon_{+;{\bf k}}$ fermion branch. 
Let us try to understand them,
\begin{align} 
\epsilon_{{\bf k};+} = \frac{1+\alpha}{2}\xi_{\bf k} + \sqrt{\left( \frac{1-\alpha}{2}\xi_{\bf k}  \right)^2 + \bar{\theta}\theta k^2} = 0,
\end{align}
which can be solved only when $\xi_{\bf k} < 0$. Then, the equation is rewritten as
\begin{align}
\alpha \xi_{\bf k}^2 = \bar{\theta}\theta k^2
\end{align}

 For, $k_{y}^2 > \zeta 2m \mu$, i.e. the momentum $k_{y}$ is between the small and large pockets (forbidden zone), there is an expected edge state from the $\epsilon_{+;{\bf k}}$ band, which decays from the boundary as 
\begin{align}
\psi({\bf r}) \propto e^{-\sqrt{k_{y}^2 - \zeta 2m \mu}x} e^{ik_{y}y}.
\end{align}

\end{widetext}

\end{document}